\documentclass[twocolumn,english,superscriptaddress]{revtex4-2}
\usepackage[T1]{fontenc}
\usepackage[utf8]{inputenc}
\usepackage{xcolor}
\usepackage{babel}
\usepackage{bm}
\usepackage{amsmath}
\usepackage{graphicx}
\usepackage[bookmarks=true,bookmarksnumbered=false,bookmarksopen=false,
 breaklinks=false,pdfborder={0 0 1},backref=false,colorlinks=true]
 {hyperref}
\hypersetup{
 allcolors=blue}

\makeatletter
\usepackage{xstring}
\let\oldhref\href
\renewcommand{\href}[2]{%
    \IfBeginWith{#1}{mailto:}{%
        \begingroup
        \hypersetup{urlcolor=black}
        \oldhref{#1}{#2}%
        \endgroup
    }{%
        \oldhref{#1}{#2}%
    }%
}
\usepackage{subfig}
\usepackage{caption}
\usepackage{ragged2e}
\ifdefined\showcaptionsetup
 \PassOptionsToPackage{caption=false}{subfig}
\fi
\usepackage{subfig}
\makeatother

\begin{document}
\title{Twist-controlled thermodynamic anomalies in bilayer $q$-state clock
models}
\author{Ze-Heng Yu}
\affiliation{Institute for Theoretical Physics, School of Physics, South China
Normal University, Guangzhou 510006, China}
\affiliation{Key Laboratory of Atomic and Subatomic Structure and Quantum Control
(Ministry of Education), Guangdong Basic Research Center of Excellence
for Structure and Fundamental Interactions of Matter, School of Physics,
South China Normal University, Guangzhou 510006, China}
\affiliation{Guangdong Provincial Key Laboratory of Quantum Engineering and Quantum
Materials, Guangdong-Hong Kong Joint Laboratory of Quantum Matter,
South China Normal University, Guangzhou 510006, China}
\author{Liang He}
\email{liang.he@scnu.edu.cn}

\affiliation{Institute for Theoretical Physics, School of Physics, South China
Normal University, Guangzhou 510006, China}
\affiliation{Key Laboratory of Atomic and Subatomic Structure and Quantum Control
(Ministry of Education), Guangdong Basic Research Center of Excellence
for Structure and Fundamental Interactions of Matter, School of Physics,
South China Normal University, Guangzhou 510006, China}
\affiliation{Guangdong Provincial Key Laboratory of Quantum Engineering and Quantum
Materials, Guangdong-Hong Kong Joint Laboratory of Quantum Matter,
South China Normal University, Guangzhou 510006, China}
\begin{abstract}
Twist has emerged as a geometric means of controlling collective behavior
across a wide range of physical systems. Whether it can qualitatively
reorganize the thermodynamic anomalies of discrete spin models, and
how such changes relate to microscopic spin configurations, remains
less explored. Here we demonstrate twist-controlled thermodynamic
anomalies in ferromagnetic bilayer $q$-state clock models using Monte
Carlo simulations. With sufficiently strong interlayer coupling, twisting
changes the number and structure of specific-heat peaks: a single
peak can split into two, or two peaks can merge into one, depending
on the number of allowed spin orientations. These changes are nonmonotonic
in twist angle and occur over narrower parameter ranges, or are absent,
for the larger $q$ values examined. The accompanying spin textures
and vortex distributions suggest a physical picture in which moir\'e
modulation of the interlayer coupling reorganizes local spin alignment
and topological defects. In a representative case, peak splitting
coincides with nearly uniform spin regions arranged on the moir\'e
scale and a reduced vortex population. These findings identify twist
as a control parameter for the collective thermal response of discrete
spin systems and raise the possibility of modified phase-transition
behavior, including the emergence or suppression of an intermediate
Berezinskii--Kosterlitz--Thouless phase.
\end{abstract}
\maketitle

\section{INTRODUCTION}

Twist has emerged as a geometric means of controlling collective behavior
across a wide range of physical systems \citep{Torre_Small_2025,Chen_APE_2026,Huynh_2DM_2026}.
A relative rotation between two lattices produces a moir\'e pattern,
introducing spatial variations in local stacking and interlayer coupling
that can reshape the balance between competing energy scales. The
discoveries of correlated insulating states and unconventional superconductivity
in twisted bilayer graphene established the potential of this approach
for engineering electronic phases \citep{Cao_Nature_2018_1,Cao_Nature_2018_2,Andrei_NM_2020,Balents_NP_2020}.
Related developments in transition-metal dichalcogenides have revealed
correlated insulating states and diverse forms of charge ordering
\citep{Xu_Nature_2020,Wang_NM_2020}. In magnetic systems, spatially
varying interlayer exchange can organize ferromagnetic and antiferromagnetic
domains within a moir\'e pattern \citep{Song_Science_2021,Li_NC_2024,Mondal_NC_2026,AKram_NL_2021}.
These findings connect geometric structure to collective ordering
and phase transitions, raising the broader question of how twisting
influences the thermal behavior of interacting systems \citep{Wang_NM_2020,Song_Science_2021}.

Ultracold atoms in optical lattices provide a particularly promising
setting for exploring this connection, owing to their controllable
geometry, interactions, and coupling between internal states \citep{Bloch_RMP_2008,Gross_Science_2017,Schafer_NRP_2020,Asteria_JOPB_2025,Zhang_WS_2015}.
Twisted bilayer optical lattices have been realized using two atomic
spin states that experience independently oriented lattice potentials,
with microwave coupling providing a tunable connection between the
layers \citep{Meng_Nature_2023,Luo_PRL_2021,ZHANG_APS_2026}. Experiments
have demonstrated a superfluid--Mott-insulator transition in this
setting \citep{Meng_Nature_2023}. Theoretical studies further predict
that the interplay between twist and interactions can produce spatially
structured superfluid and insulating states, including phases arising
from interaction-induced moir\'e modulation \citep{Zeng_PRA_2025,Johnstone_PRR_2024}.
These advances extend the study of moir\'e physics to systems in
which geometry and interlayer coupling can be varied systematically.
They also make it timely to investigate how spatially modulated interactions
affect local ordering and thermal fluctuations in simple spin models
\citep{Meng_Nature_2023,Zeng_PRA_2025,Johnstone_PRR_2024}.

The $q$-state clock model offers a natural framework for addressing
this question because it connects discrete spin symmetry with the
physics of continuous planar rotations \citep{Kosterlitz_JOPC_1973,Kosterlitz_JOPC_1974,Jose_PRB_1977,Elitzur_PRD_1979}.
Each spin is restricted to $q$ equally spaced orientations, and the
conventional two-dimensional ferromagnetic model exhibits distinct
phase-transition scenarios as $q$ varies. For $q=2,3,4$, a single
continuous transition separates the ferromagnetic and paramagnetic
phases \citep{Jose_PRB_1977,Baek_PRE_2010,Li_PRE_2020}. For finite
integer $q\ge5$, two Berezinskii--Kosterlitz--Thouless (BKT) transitions
enclose an intermediate phase with quasi-long-range order \citep{Chen_PRE_2022,Li_PRE_2020,Surungan_JOPA_2019,Miyajima_PRB_2021}.
These regimes typically exhibit single- and double-peak specific-heat
structures, respectively, although the peaks do not directly locate
the BKT transitions \citep{Miyajima_PRB_2021,Negrete_Entropy_2018,Samlodia_PRD_2024}.
This association makes thermodynamic anomalies a useful starting point
for exploring possible changes in thermal ordering. A central question
is whether twisting can qualitatively reshape these anomalies in coupled
clock-model layers, and how such changes relate to the underlying
spin textures and topological defects \citep{Li_PRE_2020,Miyajima_PRB_2021}.

Here we demonstrate twist-controlled thermodynamic anomalies in ferromagnetic
bilayer $q$-state clock models using Monte Carlo simulations. With
sufficiently strong interlayer coupling, twisting can split a single
specific-heat peak into two or merge two peaks into one, with a nonmonotonic
dependence on twist angle and a sensitivity to the number of allowed
spin orientations. The accompanying spin configurations and vortex
distributions suggest a physical picture in which moir\'e modulation
of the coupling network reorganizes local spin alignment and topological
defects. In a representative case, the emergence of a double-peak
structure coincides with nearly uniform spin regions arranged on the
moir\'e scale and a reduced vortex population. These findings connect
geometric control of microscopic spin configurations to qualitative
changes in thermodynamic response. They also raise the possibility
of twist-induced modifications to the underlying phase transitions,
including the emergence or suppression of an intermediate BKT phase.

\section{SYSTEM AND MODEL}

\begin{figure}
\includegraphics[width=0.93\columnwidth]{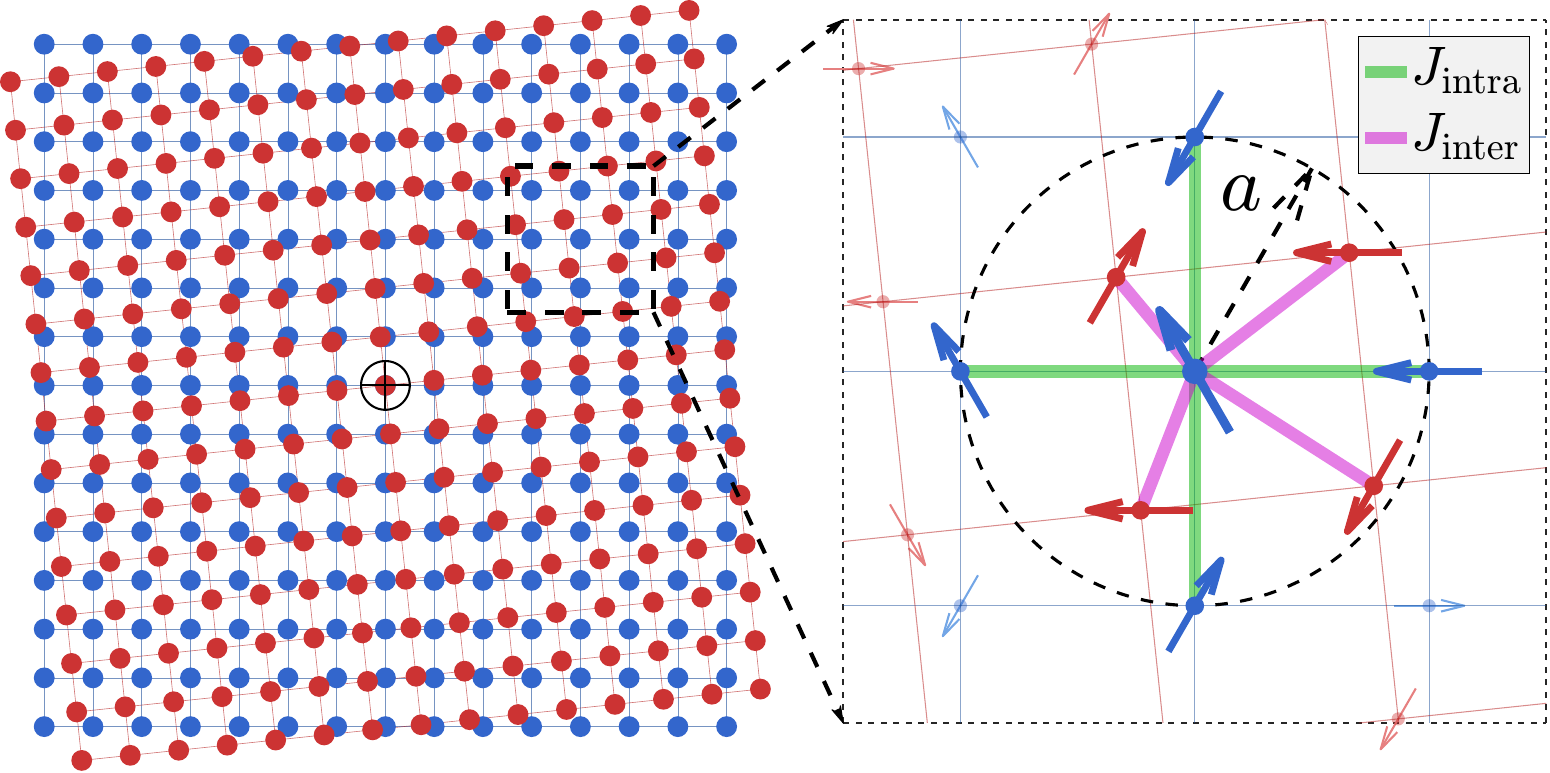}

\caption{Schematic illustration of the twisted bilayer $q$-state clock model.
Left: Two square lattices with a relative twist angle of $\Theta=6^{\circ}$.
Red (blue) dots denote lattice sites in the $\uparrow$($\downarrow$)
layer. Right: Enlarged view illustrating the spin orientations and
couplings. Red (blue) arrows represent spins in the $\uparrow$($\downarrow$)
layer, each restricted to one of $q$ discrete in-plane orientations.
Green bonds indicate nearest-neighbor intralayer couplings of strength
$J_{\mathrm{intra}}$. Purple bonds indicate interlayer couplings
of strength $J_{\mathrm{inter}}$ between sites whose in-plane separation
is within one lattice constant $a$. See the main text for details.}\label{fig:schematic_twisted_bilayer_lattice}
\end{figure}

We consider a twisted bilayer $q$-state clock model described by
the Hamiltonian
\begin{align}
H & =-J_{\mathrm{intra}}\sum_{\sigma=\uparrow,\downarrow}\sum_{\left\langle i_{\sigma},j_{\sigma}\right\rangle }\cos\left(\theta_{i_{\sigma}}-\theta_{j_{\sigma}}\right)\nonumber \\
 & \quad-J_{\mathrm{inter}}\sum_{\left\langle i_{\uparrow},j_{\downarrow}\right\rangle }\cos\left(\theta_{i_{\uparrow}}-\theta_{j_{\downarrow}}\right).
\end{align}
Here, $\sigma=\uparrow,\downarrow$ labels the two layers, each comprising
an $L\times L$ square lattice with lattice constant $a$. The spin
orientation at site $i$ in layer $\sigma$ takes one of $q$ discrete
values, $\theta_{i_{\sigma}}=2\pi n/q$, with $n=0,1,\ldots,q-1$.
The first term of $H$ describes nearest-neighbor intralayer interactions
of strength $J_{\mathrm{intra}}$, while the second describes interlayer
interactions of strength $J_{\mathrm{inter}}$. The interlayer sum
runs over pairs of sites in opposite layers whose in-plane separation
is within one lattice constant $a$. Both couplings are ferromagnetic.
The relative rotation of the two lattices is specified by the twist
angle $\Theta$, as illustrated in Fig.~\ref{fig:schematic_twisted_bilayer_lattice}.

At $J_{\mathrm{inter}}=0$, the Hamiltonian reduces to two independent
conventional square-lattice $q$-state clock models. This decoupled
limit provides a reference for understanding the bilayer system, particularly
in the weak-interlayer-coupling regime $J_{\mathrm{inter}}\ll J_{\mathrm{intra}}$.
The conventional model exhibits distinct thermal behavior depending
on $q$ \citep{Negrete_Entropy_2018,Li_PRE_2020,Miyajima_PRB_2021}.
For $q=2,3,4$, it undergoes a single continuous transition between
a low-temperature ferromagnetic phase with long-range order and a
high-temperature disordered paramagnetic phase. The specific heat
$C_{v}(T)$ diverges at the transition in the thermodynamic limit,
while finite systems exhibit a rounded peak \citep{Onsager_PR_1944,Baek_PRE_2010,Ferdinand_PR_1969,Aguilera_Entropy_2026}.
For finite integer $q\geq5$, the conventional model instead exhibits
two BKT transitions, with an intermediate phase possessing quasi-long-range
order between the ferromagnetic and paramagnetic phases. The specific
heat typically exhibits two broad, nondivergent peaks, whose temperatures
generally differ from the BKT transition temperatures \citep{Miyajima_PRB_2021,Zhang_PLA_1990,Hostetler_PRD_2021,Nguyen_PRB_2020,Samlodia_PRD_2024}.
Although these peaks do not directly locate the transitions, the association
of distinct peak structures with different phase-transition scenarios
provides a useful reference for assessing possible changes in the
underlying transition behavior of the system.

Beyond this conventional limit, the bilayer geometry introduces twist
as an additional control parameter. Recent studies have demonstrated
that twist can modify phases and phase transitions in a variety of
physical systems \citep{Cao_Nature_2018_1,Cao_Nature_2018_2,Andrei_NM_2020,Balents_NP_2020,Xu_Nature_2020,Wang_NM_2020}.
Finite-temperature phase transitions often manifest themselves through
anomalies in thermodynamic observables, such as the specific heat
\citep{Choy_PT_1990,Gonzalez_PRB_2021}. These signatures provide
a natural starting point for investigating how twist affects ordering
at finite temperatures, even when the anomalies do not directly locate
the transitions. Motivated by this connection, we investigate how
twisting modifies the specific-heat anomalies of the bilayer $q$-state
clock model.

In the present model, twisting changes the interlayer coupling network
through the relative positions of sites in the two layers. When $J_{\mathrm{inter}}$
is sufficiently large compared with $J_{\mathrm{intra}}$, these geometric
changes can substantially modify the thermal response. As shown below,
twisting can qualitatively change the number and shape of the specific-heat
peaks (Fig.~\ref{fig:Fig_2}). Given the distinct peak structures
associated with different phase-transition scenarios in the conventional
clock model, these changes motivate further investigation of possible
modifications to the underlying phase transitions. Here, we focus
on the specific-heat anomalies of this system.

\section{RESULTS}\label{sec:RESULTS}

Most of the numerical results presented below are obtained using Monte
Carlo simulations (see Appendix \ref{sec:Simulation-details} for
computational details). Unless otherwise stated, each layer consists
of a $256\times256$ square lattice with open boundary conditions.
We focus on thermodynamic anomalies in the specific heat and assess
the effects of system size and boundary conditions. Simulations of
larger systems and calculations using alternative boundary conditions
yield no significant changes in the reported features (see Appendices
\ref{sec:Finite-size-effects} and \ref{sec:Boundary_conditions}).

\subsection{Twist-controlled thermodynamic anomalies}\label{subsec:III A}

\begin{figure}
\includegraphics[width=3.4in]{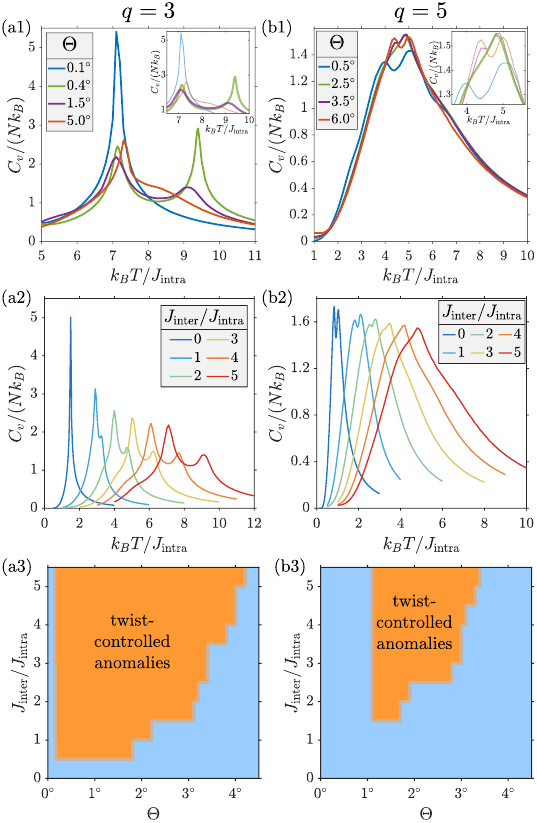}

\caption{Twist-controlled specific-heat anomalies in the bilayer 3-state (left
column) and 5-state (right column) clock models. (a1), (b1) Temperature
dependence of the specific heat $C_{v}$ for selected twist angles
$\Theta$ at a fixed interlayer coupling $J_{\mathrm{inter}}/J_{\mathrm{intra}}=5$.
The insets show enlarged views of the peak region. For $q=3$, increasing
the twist angle from ${0.1}^{\circ}$ to ${0.4}^{\circ}$ changes
the specific-heat curve from a single-peak to a double-peak structure.
For $q=5$, increasing the twist angle can instead merge the two peaks
into a single broad peak. (a2), (b2) Temperature dependence of $C_{v}$
for selected interlayer coupling strengths $J_{\mathrm{inter}}/J_{\mathrm{intra}}$
at fixed twist angles $\Theta={1.5}^{\circ}$ and ${2.5}^{\circ}$,
respectively. (a3), (b3) Regions highlighted in orange in the $\Theta-J_{\mathrm{inter}}/J_{\mathrm{intra}}$
plane indicate where twisting produces a specific-heat peak structure
that differs qualitatively from that of the corresponding single-layer
model. See the main text for details.}\label{fig:Fig_2}
\end{figure}

We begin with the $q=3$ case. Fig.~\ref{fig:Fig_2}(a1) shows the
temperature dependence of the specific heat $C_{v}(T)$ for selected
twist angles at a fixed interlayer coupling $J_{\mathrm{inter}}/J_{\mathrm{intra}}=5$.
At the small twist angle $\Theta={0.1}^{\circ}$, the specific heat
exhibits a single peak {[}blue curve in Fig.~\ref{fig:Fig_2}(a1){]},
resembling the one of the conventional single-layer 3-state clock
model. This similarity may be understood qualitatively by noting that,
over the finite system considered here, the local stacking varies
only weakly at such a small twist angle. The ferromagnetic interlayer
coupling may therefore act primarily to reinforce spin alignment,
preserving a thermal response similar to that of the conventional
model. Increasing the twist angle to $\Theta={0.4}^{\circ}$ produces
a qualitative change: the single peak develops into a double-peak
structure {[}green curve in Fig.~\ref{fig:Fig_2}(a1){]}. This structure
resembles the specific-heat response of conventional clock models
with $q\geq5$\textcolor{gray}{{} \citep{Negrete_Entropy_2018,Li_PRE_2020,Miyajima_PRB_2021}},
in which two BKT transitions enclose an intermediate phase with quasi-long-range
order. By analogy, the observed peak splitting suggests the possibility
that twisting induces an intermediate BKT phase in the bilayer 3-state
clock model. The double-peak structure persists up to approximately
$\Theta={4.2}^{\circ}$, beyond which the specific heat returns to
a single-peak form {[}see Fig.~\ref{fig:Fig_2}(a1,~a3){]}.

To examine the role of interlayer coupling, we calculate $C_{v}(T)$
for different values of $J_{\mathrm{inter}}/J_{\mathrm{intra}}$ at
a fixed twist angle $\Theta={1.5}^{\circ}$ {[}see Fig.~\ref{fig:Fig_2}(a2){]}.
A double-peak structure can emerge when the interlayer coupling is
comparable to the intralayer coupling, as illustrated by the light-blue
curve. Thus, even a small twist can qualitatively alter the specific-heat
peak structure when the interlayer interaction is sufficiently strong.
We refer to such qualitative changes relative to the corresponding
single-layer model as twist-controlled thermodynamic anomalies. Figure~\ref{fig:Fig_2}(a3)
summarizes the parameter region in which these twist-controlled anomalies
occur, highlighted in orange in the $\Theta-J_{\mathrm{inter}}/J_{\mathrm{intra}}$
plane. Similar behavior is found for $q=4$, with detailed results
presented in Appendix~\ref{sec:q=00003D4 and 6}.

For $q=5$, twisting also changes the specific-heat peak structure,
but the change occurs in the opposite direction. Figure~\ref{fig:Fig_2}(b1)
shows $C_{v}(T)$ for selected twist angles at $J_{\mathrm{inter}}/J_{\mathrm{intra}}=5$.
At $\Theta={0.5}^{\circ}$, the specific heat exhibits two peaks {[}blue
curve in Fig.~\ref{fig:Fig_2}(b1){]}, resembling the one associated
with the two BKT transitions of the conventional 5-state clock model
\citep{Negrete_Entropy_2018,Li_PRE_2020,Miyajima_PRB_2021}. As the
twist angle increases to $\Theta={2.5}^{\circ}$, the two peaks merge
into a single broad peak {[}green curve in Fig.~\ref{fig:Fig_2}(b1){]}.
This change raises the possibility that twisting suppresses the intermediate
BKT phase, potentially replacing the two-transition sequence with
a single transition between the ferromagnetic and paramagnetic phases.
At larger twist angles, the double-peak structure reappears.

Calculations at a fixed twist angle $\Theta={2.5}^{\circ}$ further
show that sufficiently strong interlayer coupling enables the merging
of the two peaks {[}see Fig.~\ref{fig:Fig_2}(b2){]}. The corresponding
region of twist-controlled thermodynamic anomalies is shown in Fig.~\ref{fig:Fig_2}(b3).
Compared with the $q=3$ case {[}see Fig.~\ref{fig:Fig_2}(a3){]},
the region of twist-controlled thermodynamic anomalies for $q=5$
{[}Fig.~\ref{fig:Fig_2}(b3){]} occupies a smaller portion of the
parameter plane. This narrowing suggests that the influence of twist
on the qualitative specific-heat peak structure may become less pronounced
as $q$ increases.

For the values of $q\ge6$ examined, no qualitative change in the
specific-heat peak structure is observed within the investigated ranges,
$\Theta\in[{0}^{\circ},{6}^{\circ}]$ and $J_{\mathrm{inter}}/J_{\mathrm{intra}}\in[0,5]$
(see Appendix~\ref{sec:q=00003D4 and 6}). A possible qualitative
explanation is that, as $q$ increases, the more closely spaced allowed
spin orientations provide greater flexibility for local spin configurations
to adjust to the spatially varying interlayer coupling network. This
flexibility may allow the system to accommodate twist-induced geometric
changes without qualitatively altering the specific-heat peak structure.

The possible phase-transition scenarios discussed above for $q=3$
and $q=5$ are inferred by analogy with the specific-heat behavior
of conventional clock models. These interpretations remain tentative:
changes in the number of peaks alone do not establish changes in the
number or nature of the phase transitions in the twisted bilayer system.
To gain further insight into the microscopic behavior accompanying
these twist-controlled anomalies, we next examine the real-space spin
configurations and vortex distributions.

\subsection{Twist-controlled vortex structure and thermodynamic anomalies}

\begin{figure*}
\includegraphics[width=6.9in]{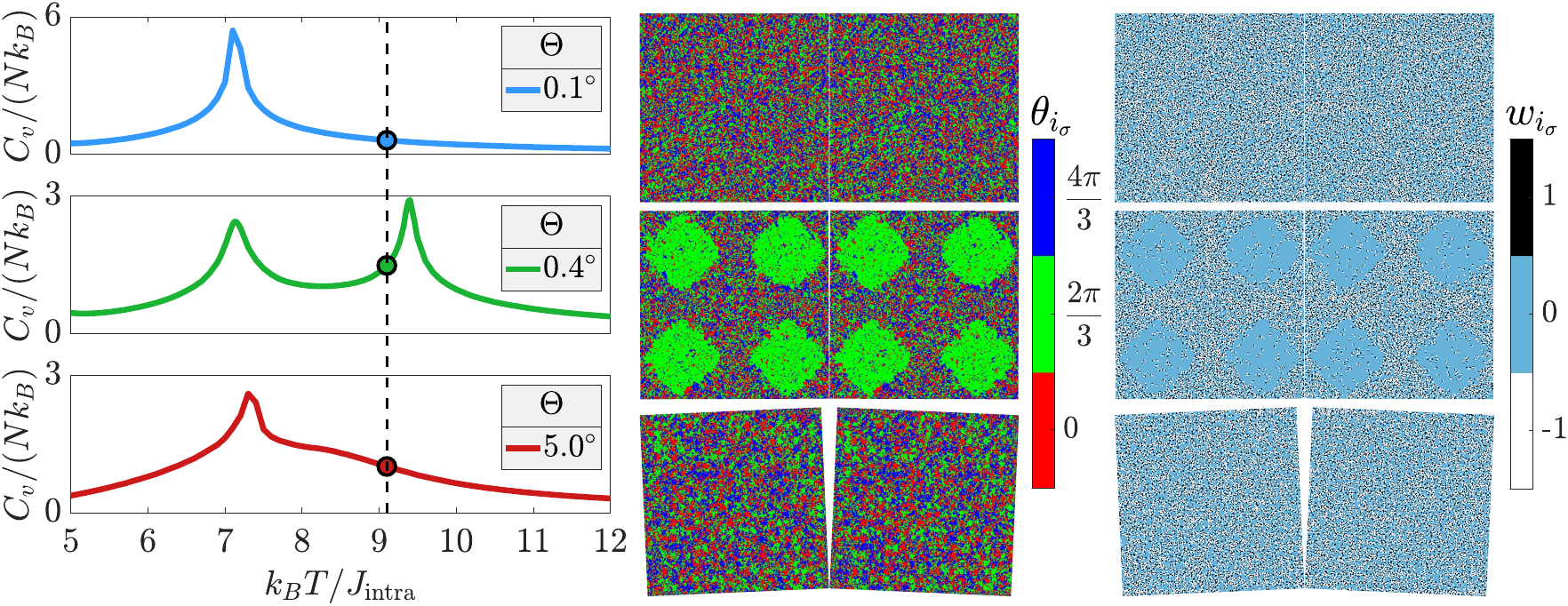}

\caption{Specific heat, representative spin configurations, and local winding-number
distributions of the twisted bilayer 3-state clock model at different
twist angles $\Theta$, with $J_{\mathrm{inter}}/J_{\mathrm{intra}}=5$.
Left: Temperature dependence of the specific heat $C_{v}$ for $\Theta={0.1}^{\circ}$
(blue), ${0.4}^{\circ}$ (green), and ${5.0}^{\circ}$ (red). The
dashed line and circles mark the common temperature $k_{B}T/J_{\mathrm{intra}}=9.1$
at which the configurations in the middle and right columns are sampled.
Middle: Real-space spin configurations, with colors indicating the
local spin orientation $\theta_{i_{\sigma}}$. Right: Corresponding
local winding-number distributions, where $w_{i_{\sigma}}=+1$ and
$-1$ identify vortices and antivortices, respectively, and the blue
background denotes $w_{i_{\sigma}}=0$. The combined numbers of vortices
and antivortices in the displayed configurations are $2.5\times10^{4}$,
$1.7\times10^{4}$, and $2.2\times10^{4}$, from top to bottom, respectively.
As $\Theta$ increases from ${0.1}^{\circ}$ to ${0.4}^{\circ}$,
large regions of nearly uniform spin orientation emerge and the combined
vortex and antivortex count decreases. These changes accompany the
emergence of a double-peak structure in the specific heat. At ${5.0}^{\circ}$,
the specific heat again exhibits a single peak, while the spin configurations
show smaller regions of similar orientation and a higher combined
vortex and antivortex count than at ${0.4}^{\circ}$. See the main
text for details.}\label{fig:Fig_3}
\end{figure*}

We next examine the microscopic changes accompanying the twist-controlled
specific-heat anomalies. Of particular interest is the emergence of
a double-peak structure for $q=3,4$, which resembles the specific-heat
response of conventional clock models exhibiting an intermediate BKT
phase. This resemblance raises the possibility that twisting induces
an intermediate phase with quasi-long-range order. To explore the
microscopic behavior associated with this possibility, we examine
the real-space spin configurations and vortex distributions, using
the $q=3$ case as a representative example. Figure~\ref{fig:Fig_3}
presents the specific heat, spin configurations, and local winding-number
$w_{i_{\sigma}}\equiv[(\theta_{i_{\sigma,x}+1,i_{\sigma,y}}-\theta_{i_{\sigma,x},i_{\sigma,y}})+(\theta_{i_{\sigma,x}+1,i_{\sigma,y}+1}-\theta_{i_{\sigma,x}+1,i_{\sigma,y}})+(\theta_{i_{\sigma,x},i_{\sigma,y}+1}-\theta_{i_{\sigma,x}+1,i_{\sigma,y}+1})+(\theta_{i_{\sigma,x},i_{\sigma,y}}-\theta_{i_{\sigma,x},i_{\sigma,y}+1})]/2\pi$
distributions of the twisted bilayer 3-state clock model at $J_{\mathrm{inter}}/J_{\mathrm{intra}}=5$.

We compare configurations at the common temperature $k_{B}T/J_{\mathrm{intra}}=9.1$,
indicated by the dashed line and circles in the left column of Fig.~\ref{fig:Fig_3}.
At $\Theta={0.1}^{\circ}$ and $\Theta={5.0}^{\circ}$, this temperature
lies above the single specific-heat peak, whereas at $\Theta={0.4}^{\circ}$,
it lies between the two peaks. Increasing $\Theta$ from ${0.1}^{\circ}$
to ${0.4}^{\circ}$ changes the spin configuration from an apparently
disordered pattern to one containing large regions of nearly uniform
spin orientation. This change is accompanied by a pronounced reduction
in the total vortex count, from approximately $2.5\times10^{4}$ to
$1.7\times10^{4}$ in the displayed configurations {[}middle and right
columns of Fig.~\ref{fig:Fig_3}{]}. As $\Theta$ increases further
to ${5.0}^{\circ}$, the large, nearly uniform regions disappear,
and the total vortex count rises to approximately $2.2\times10^{4}$.
The spatial arrangement of the nearly uniform regions at $\Theta={0.4}^{\circ}$
can be understood qualitatively in terms of the moir\'e length scale.
For small twist angles, the moir\'e period is approximately $\lambda\approx a/\Theta$.
At $\Theta={0.4}^{\circ}$, each layer spans roughly two moir\'e
periods along each direction. This estimate is consistent with the
approximately $2\times2$ arrangement of nearly uniform spin regions
observed in each layer {[}middle row of Fig.~\ref{fig:Fig_3}{]}.
At $\Theta={0.1}^{\circ}$, by contrast, the moir\'e period exceeds
the system size, whereas at $\Theta={5.0}^{\circ}$, it is much smaller,
which may limit the spatial extent of regions with nearly uniform
spin orientation. These different geometric scales provide a qualitative
picture of why prominent, spatially separated regions of nearly uniform
spin orientation appear at the intermediate twist angle.

The emergence of these regions and the reduction in the total vortex
count at $\Theta={0.4}^{\circ}$ suggest enhanced spatial spin correlations.
Their concurrence with the double-peak specific-heat structure connects
the thermodynamic anomalies to a reorganization of the spin and vortex
configurations. Taken together, these observations are consistent
with the possibility, discussed in Sec.~\ref{subsec:III A}, that
twisting induces an intermediate BKT phase. Establishing this interpretation,
however, requires direct evidence of quasi-long-range order and an
appropriate finite-size analysis, which are beyond the scope of the
present work.

\section{CONCLUSIONS}

Our results reveal twist-controlled thermodynamic anomalies in the
bilayer $q$-state clock model: sufficiently strong interlayer coupling
allows twisting to qualitatively reshape the specific-heat peak structure.
The splitting and merging of peaks reveal a sensitivity to moir\'e
geometry that depends on the discreteness of the spin orientations.
The accompanying spin textures and vortex distributions suggest a
physical picture in which the spatial modulation of interlayer coupling
reorganizes local spin alignment and topological defects. These findings
establish twist as a geometric control parameter for thermodynamic
anomalies and raise the possibility of modifying the underlying phase
transitions. A natural next step is to determine whether the observed
reorganization supports an intermediate phase with quasi-long-range
order.
\begin{acknowledgments}
This work was supported by the NKRDPC (Grant No.~2022YFA1405304),
NSFC (Grant No.~12275089), and Guangdong Provincial Quantum Science
Strategic Initiative (Grant Nos.~GDZX2304002 and GDZX2404001).
\end{acknowledgments}

\appendix

\section{Simulation details}\label{sec:Simulation-details}

The Monte Carlo simulations use local Metropolis updates \citep{Metropolis_TJOCP_1953}
and start from random spin configurations. The thermodynamic quantities
in this paper are obtained as ensemble averages over $N_{\mathrm{MCS}}=1\times10^{6}$
Monte Carlo steps. The real-space spin configurations in Fig.~\ref{fig:Fig_3}
are snapshots taken after $N_{\mathrm{MCS}}$ Monte Carlo steps. Unless
otherwise stated, each layer consists of a $256\times256$ square
lattice with open boundary conditions (OBC), and the interlayer coupling
is fixed at $J_{\mathrm{inter}}/J_{\mathrm{intra}}=5$. All other
system parameters are listed in the corresponding figure captions.

In the $q$-state clock model, each spin $\bm{s}_{i}$ is confined
to rotate in a plane and can be represented by a complex phase factor
$\bm{s}_{i}=e^{\mathrm{i}\theta_{i}}$, where $\mathrm{i}$ is the
imaginary unit and $\theta_{i}$ denotes the spin orientation at site
$i$. The complex magnetization of layer $\sigma$ is defined as $\bm{M}_{\sigma}=N^{-1}_{\sigma}\sum^{N_{\sigma}}_{i=1}\bm{s}_{i_{\sigma}}=M_{\sigma}e^{\mathrm{i}\phi_{\sigma}}$,
where $N_{\sigma}$ is the number of sites in layer $\sigma$, $M_{\sigma}=\left|\bm{M}_{\sigma}\right|$
is its magnitude, and $\phi_{\sigma}$ its average phase.

The total complex magnetization of the bilayer system is $\bm{M}=(N_{\uparrow}\bm{M}_{\uparrow}+N_{\downarrow}\bm{M}_{\downarrow})/(N_{\uparrow}+N_{\downarrow})$.
For a symmetric system with $N_{\uparrow}=N_{\downarrow}=N_{\mathrm{layer}}$
and total number of sites $N=2N_{\mathrm{layer}}$, this simplifies
to $\bm{M}=N^{-1}\sum_{\sigma=\uparrow,\downarrow}\sum^{N_{\mathrm{layer}}}_{i=1}e^{\mathrm{i}\theta_{i_{\sigma}}}$,
whose magnitude is $\left|\bm{M}\right|=N^{-1}\left|\sum_{\sigma=\uparrow,\downarrow}\sum^{N_{\mathrm{layer}}}_{i=1}e^{\mathrm{i}\theta_{i_{\sigma}}}\right|$.
The layer-averaged magnetization magnitude $\left|\bm{M}\right|_{s}$
quantifies the intralayer order independently of interlayer phase
coherence: $\left|\bm{M}\right|_{s}=(N_{\uparrow}\left|\bm{M}_{\uparrow}\right|+N_{\downarrow}\left|\bm{M}_{\downarrow}\right|)/(N_{\uparrow}+N_{\downarrow})$.
For a symmetric system this reduces to $\left|\bm{M}\right|_{s}=N^{-1}\sum_{\sigma=\uparrow,\downarrow}\left|\sum^{N_{\mathrm{layer}}}_{i=1}e^{\mathrm{i}\theta_{i_{\sigma}}}\right|$.
The total magnetization magnitude $\left\langle \left|\bm{M}\right|\right\rangle $
captures global magnetic order across both layers and is estimated
from Monte Carlo data as $\left\langle \left|\bm{M}\right|\right\rangle \approx N^{-1}_{\text{MCS}}\sum^{N_{\text{MCS}}}_{k=1}\left|\bm{M}\right|^{(k)}$,
where $\left|\bm{M}\right|^{(k)}$ denotes the instantaneous value
of $\left|\bm{M}\right|$ at the $k$-th Monte Carlo step. Similarly,
the intralayer order is given by the Monte Carlo average of the layer-averaged
magnetization: $\left\langle \left|\bm{M}\right|_{s}\right\rangle \approx N^{-1}_{\text{MCS}}\sum^{N_{\text{MCS}}}_{k=1}\left|\bm{M}\right|^{(k)}_{s}$,
where $\left|\bm{M}\right|^{(k)}_{s}$ the instantaneous value of
$\left|\bm{M}\right|_{s}$ at step $k$.

The energy of the $k$-th Monte Carlo configuration $\{\theta_{i}\}^{(k)}$
is given by $E^{(k)}=H(\{\theta_{i}\}^{(k)})$. The specific heat,
which probes energy fluctuations and phase transitions, is defined
as $C_{v}=(\langle H^{2}\rangle-\langle H\rangle^{2})/(k_{B}T^{2})$,
where $\langle H^{n}\rangle\approx N^{-1}_{\mathrm{MCS}}\sum^{N_{\mathrm{MCS}}}_{k=1}(E^{(k)})^{n}$
denotes the $n$-th moment of the energy, $k_{B}$ is Boltzmann's
constant, and $T$ is the temperature.

\section{Temperature dependence of the vortex number}

\begin{figure}
\centering
\subfloat[$q=3$]{\includegraphics[width=0.45\columnwidth]{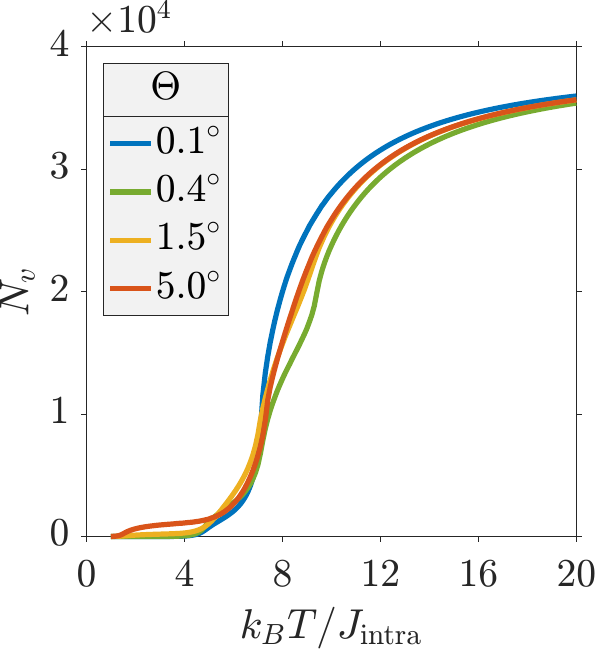}

}\subfloat[$q=5$]{\includegraphics[width=0.45\columnwidth]{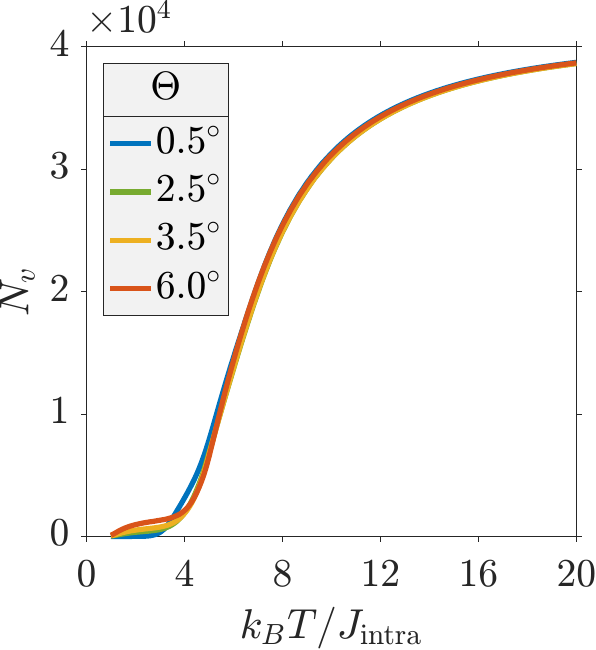}

}

\caption{Total vortex number $N_{v}$ on the $\uparrow$ and $\downarrow$
layers versus temperature $T$ for the twisted bilayer $q$-state
clock model. Panels (a) and (b) correspond to $q=3$ and 5, respectively,
with each panel showing four twist angles. In both panels, the curves
for different $\Theta$ converge to a common value in the low- and
high-temperature limits. For $q=3$, differences among the curves
appear mainly in the intermediate-temperature range.}\label{fig:Nv_T}
\end{figure}

As an additional diagnostic of the anomalous thermodynamic response,
we examine the dependence of vortex number $N_{v}$ on temperature
$T$. We consider $q=3$ and 5, and select twist angles matching those
in Fig.~\ref{fig:Fig_2}(a1,~b1), respectively.

Fig.~\ref{fig:Nv_T} presents $N_{v}$ versus $T$ for different
$q$ and $\Theta$. In both panels, the curves approach common values
in the low- and high-temperature limits, consistent with similar thermodynamic
behavior across twist angles in these regimes. For $q=3$, a pronounced
splitting occurs primarily in the intermediate-temperature range,
where the specific-heat curves exhibit their anomalous behavior. This
correlation suggests that the anomaly in $C_{v}$ is tied to a modification
of the vortex density \citep{Nelson_PRB_1979,Young_PRB_1979}. Under
open boundary conditions (OBC), topological charges are created and
annihilated at the boundaries, leading to a small instantaneous imbalance
between vortex and antivortex numbers. In the ensemble average, this
imbalance is small; the vortex and antivortex curves are symmetric
within our numerical resolution and approximately equal to one-half
of $N_{v}$, and are therefore omitted for brevity.

\section{Finite-size effects}\label{sec:Finite-size-effects}

\begin{figure}
\centering
\subfloat[$q=3,\Theta={1.5}^{\circ}$]{\includegraphics[width=0.49\columnwidth]{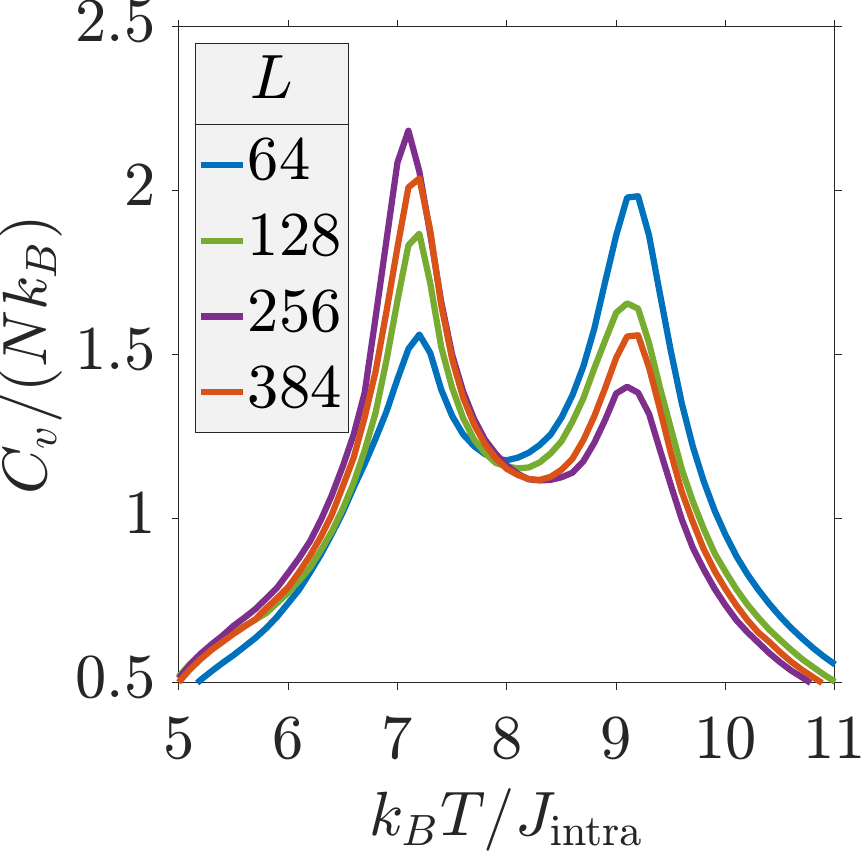}

}\subfloat[$q=5,\Theta={2.5}^{\circ}$]{\includegraphics[width=0.483\columnwidth]{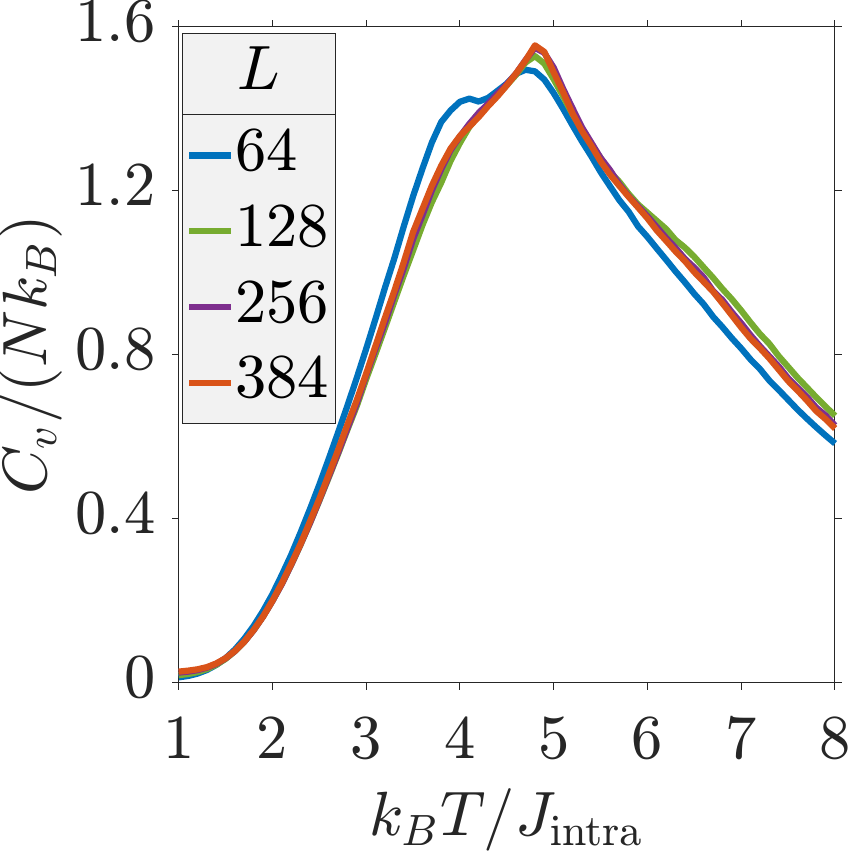}

}

\caption{Finite-size effects in the twisted bilayer $q$-state clock model.
Panels (a) and (b) show the specific-heat curves for $q=3$ at $\Theta=1.5^{\circ}$
and $q=5$ at $\Theta=2.5^{\circ}$, respectively. Each panel contains
curves for linear system sizes $L=64,128,256$, and 384. As $L$ increases,
the specific-heat profiles evolve while retaining the anomalous peak
features characteristic of each $q$.}\label{fig:ApC}
\end{figure}

To further demonstrate that the anomalous response in the specific-heat
curves of twisted bilayer clock models persists across different system
sizes, we perform Monte Carlo simulations for $L=64,128,256,$ and
384.

As shown in Fig.~\ref{fig:ApC}, each panel contains four specific
heat $C_{v}$ versus temperature $T$ curves, corresponding to four
different system sizes $L$. The $L=256$ curve serves as the benchmark,
as this is the standard system size used in the main text. The curves
for $L=64,128,$ and 384---whose linear dimensions are 1/4, 1/2,
and 1.5 times that of $L=256$, respectively---are used to investigate
the thermodynamic response in smaller and larger systems. As shown
in Fig.~\ref{fig:ApC}(a) for the $q=3$ clock model, the double-peak
anomaly appears in the specific heat curves for all four system sizes,
confirming that this thermodynamic response is robust across different
$L$. A closer comparison of these curves reveals that at $L=64$,
the left peak is lower than the right peak. As $L$ increases, the
left peak gradually grows while the right peak becomes suppressed.
This trend can be attributed to the existence of an anomalous-angle
window in the twisted bilayer clock model: with increasing $L$, this
window shifts toward smaller twist angles, so that a curve that is
anomalous at small $L$ may fall into the normal region at larger
$L$, and vice versa. As shown in Fig.~\ref{fig:ApC}(b) for the
$q=5$ clock model, the anomalous single-peak feature in the $C_{v}$
curves is also present for system sizes $L=128,256,$ and 384. In
summary, the thermodynamic anomaly is robust across different system
sizes and originates from the intrinsic bilayer geometry rather than
from finite-size effects.

\section{Results for different boundary conditions}\label{sec:Boundary_conditions}

\begin{figure}
\centering
\subfloat[$q=3$ (PBC)]{\includegraphics[width=0.48\columnwidth]{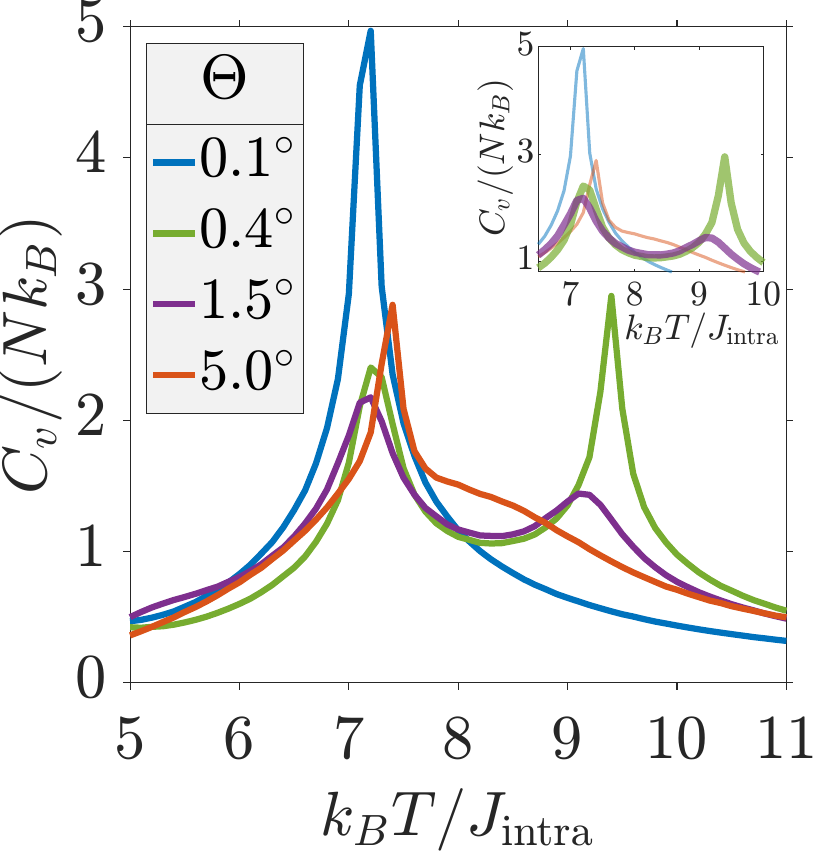}

}\subfloat[$q=3$ (CBC)]{\includegraphics[width=0.48\columnwidth]{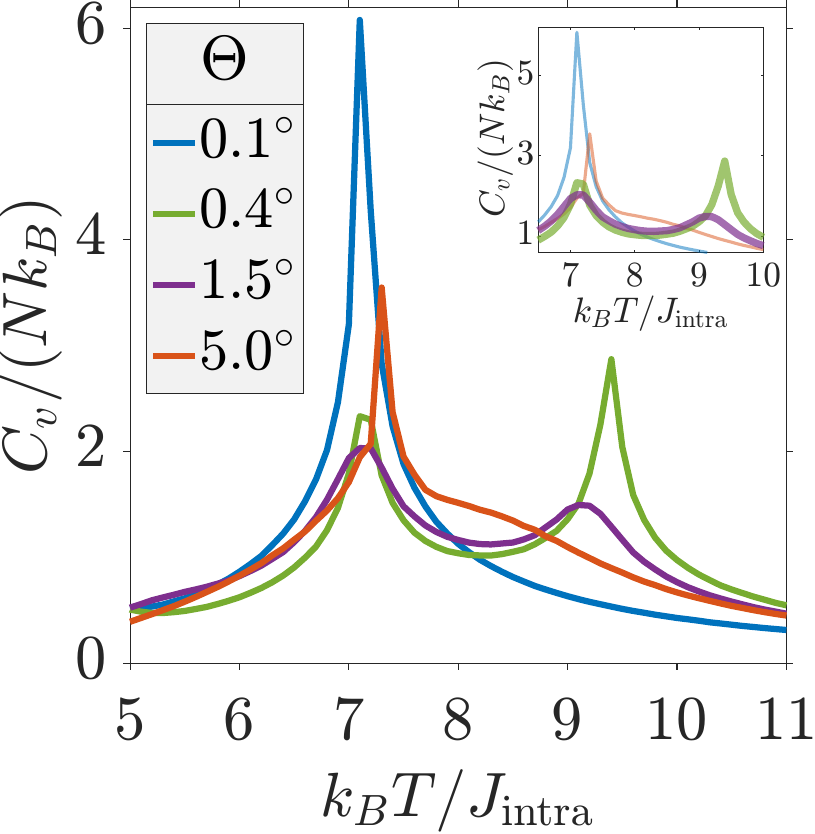}

}

\subfloat[$q=5$ (PBC)]{\includegraphics[width=0.48\columnwidth]{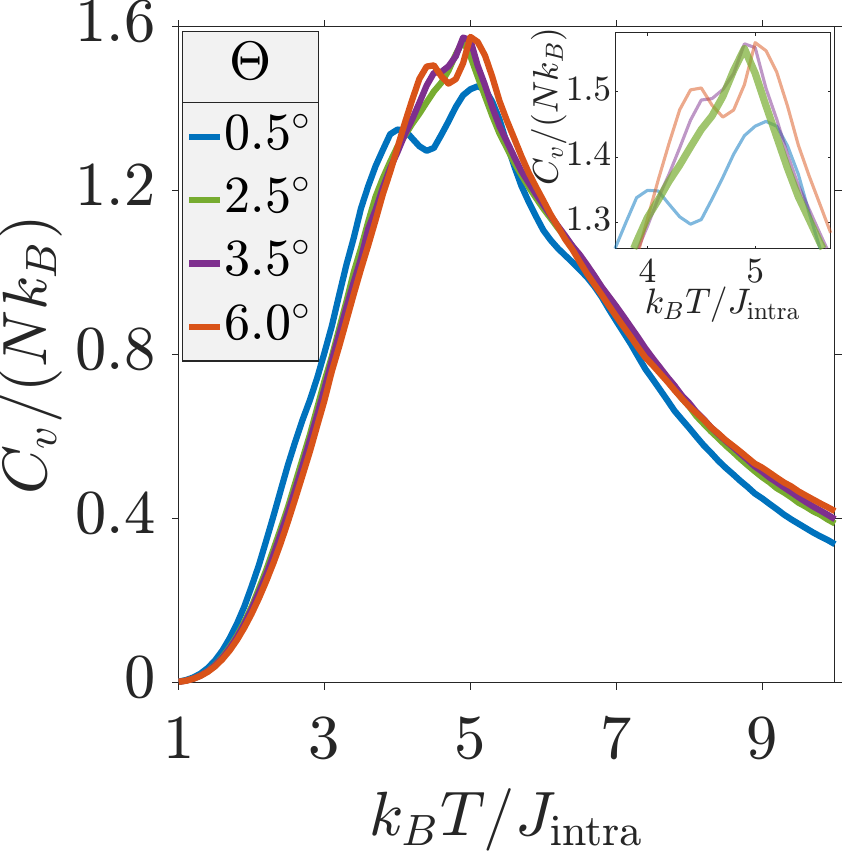}

}\subfloat[$q=5$ (CBC)]{\includegraphics[width=0.48\columnwidth]{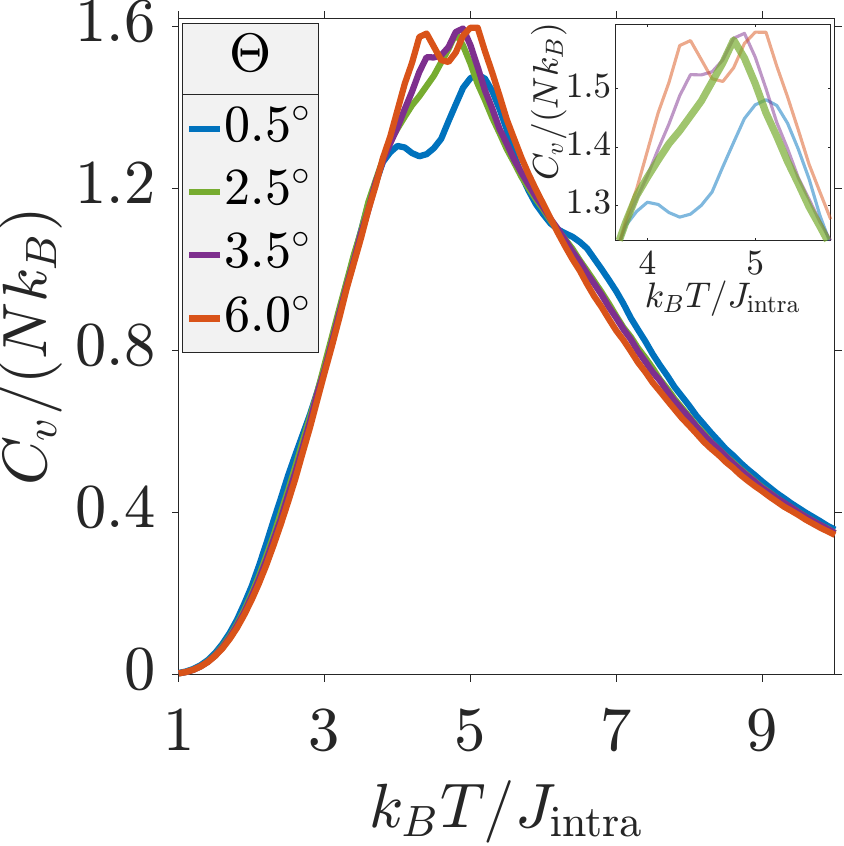}

}

\caption{Comparison of boundary conditions. Panels (a) and (b) correspond to
$q=3$, while panels (c) and (d) correspond to $q=5$. Panels (a)
and (c) are obtained under periodic boundary conditions (PBC), whereas
panels (b) and (d) are obtained under circular boundary conditions
(CBC). Each panel contains four specific-heat curves for different
twist angles $\Theta$. As $\Theta$ varies, the curves for both $q=3$
and $q=5$ exhibit single-double peak transitions under both boundary
conditions, and the resulting thermodynamic behavior remains consistent
with that observed under open boundary conditions (OBC) in the main
text.}\label{fig:ApD}
\end{figure}

The main text focuses primarily on the bilayer system with open boundary
conditions (OBC), which are also the most commonly adopted in twisted
bilayer experiments. To test whether the anomalous thermodynamic response
also occurs under other boundary conditions, we perform additional
simulations with periodic boundary conditions (PBC) and circular boundary
conditions (CBC) \citep{Joseph_JOPA_1996,Toldin_PRR_2025}. For the
PBC case, the two superimposed monolayers, each subject to its own
periodic boundary conditions, are twisted relative to one another.
For the CBC case, the spin sites are restricted to the region inside
the common inscribed circle of the two square lattices. Keeping all
other simulation parameters the same as in the OBC studies, we examine
the thermodynamic behavior under these two boundary conditions.

As shown in Fig.~\ref{fig:ApD}, which displays the simulation results
for the $q=3$ and $q=5$ clock models at various twist angles $\Theta$,
the single-double peak transition also occurs under both boundary
conditions. Specifically, under PBC and CBC, as $\Theta$ increases,
the $C_{v}$ curves of the $q=3$ model evolve from a single peak
to two distinct peaks and then back to a single peak, whereas those
of the $q=5$ model evolve from two peaks to a single peak and then
back to two peaks. These behaviors are consistent with those observed
under OBC. Moreover, for a given twist angle $\Theta$, the peak shapes
and characteristic temperatures are nearly identical across these
three boundary conditions, with only minor differences in the peak
heights. This comparison indicates that the thermodynamic anomaly
observed in the specific-heat curves is not an artifact of the boundary
conditions. Consequently, in experimental realizations, the boundary
conditions can be chosen according to the material properties, and
this phenomenon should in principle be observable under any of them.

\section{$q=4$ and $q=6$ cases}\label{sec:q=00003D4 and 6}

\begin{figure*}
\centering
\subfloat[$q=4$ (OBC)]{\includegraphics[width=0.25\textwidth]{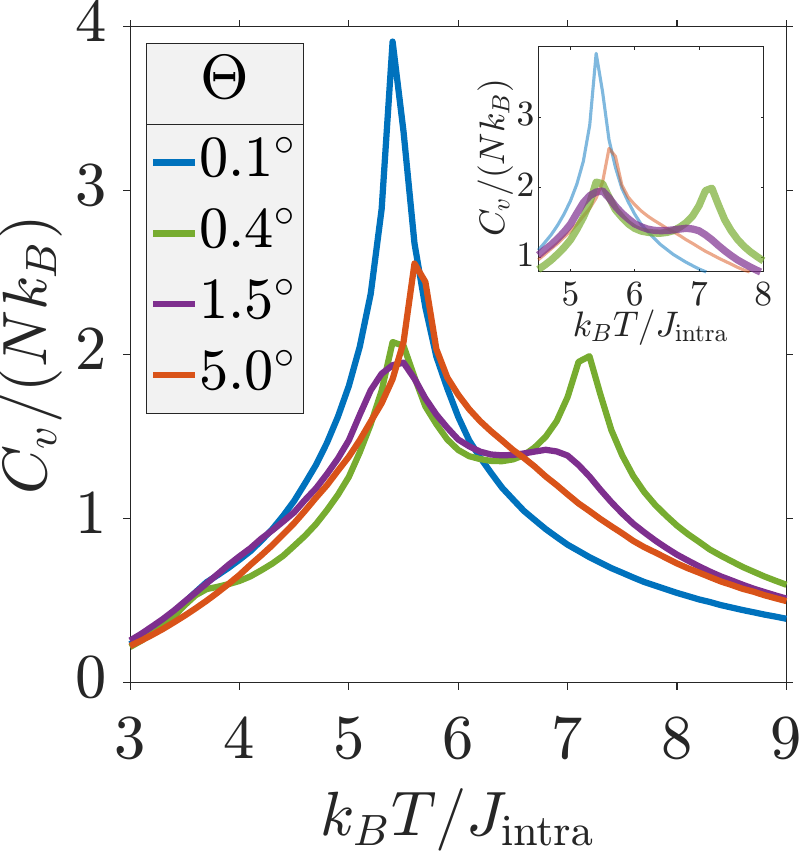}

}\subfloat[$q=4$ (PBC)]{\includegraphics[width=0.25\textwidth]{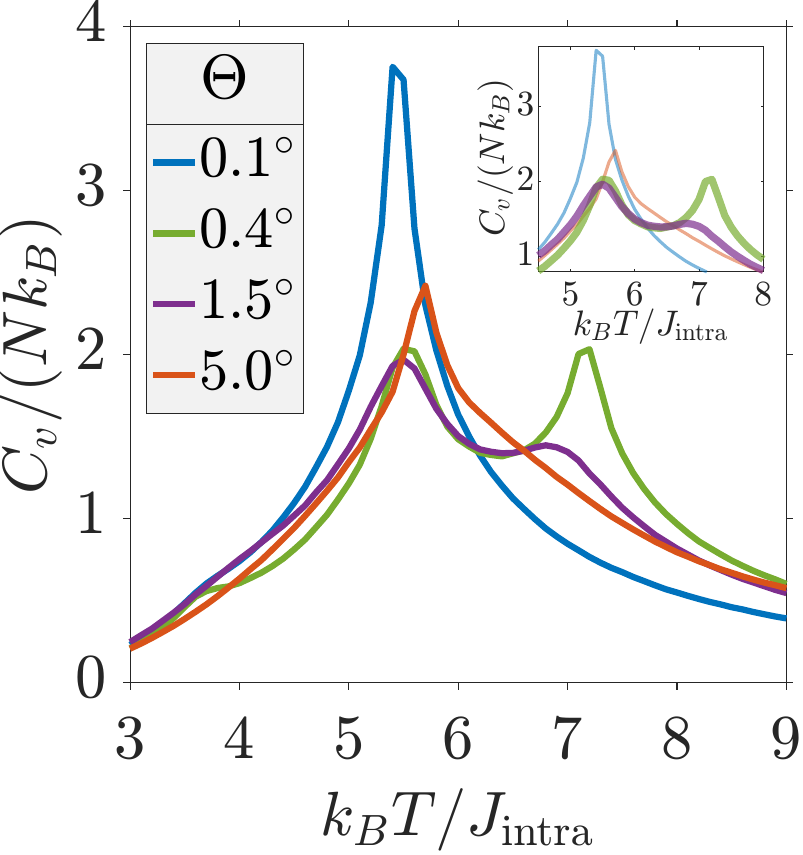}

}\subfloat[$q=4$ (CBC)]{\includegraphics[width=0.25\textwidth]{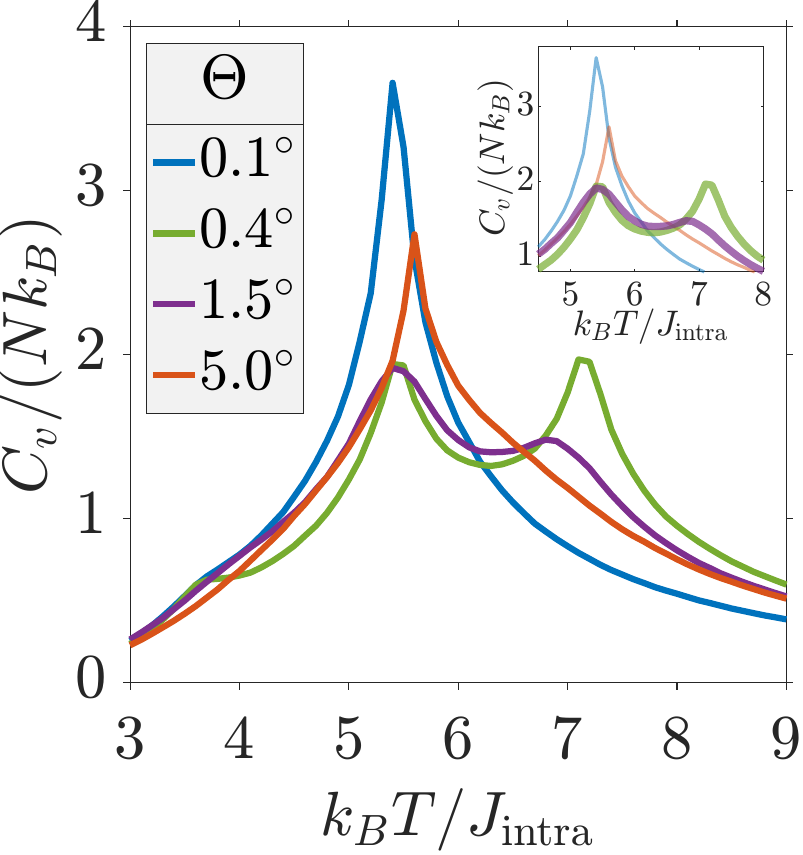}

}

\subfloat[$q=6$ (OBC)]{\includegraphics[width=0.263\textwidth]{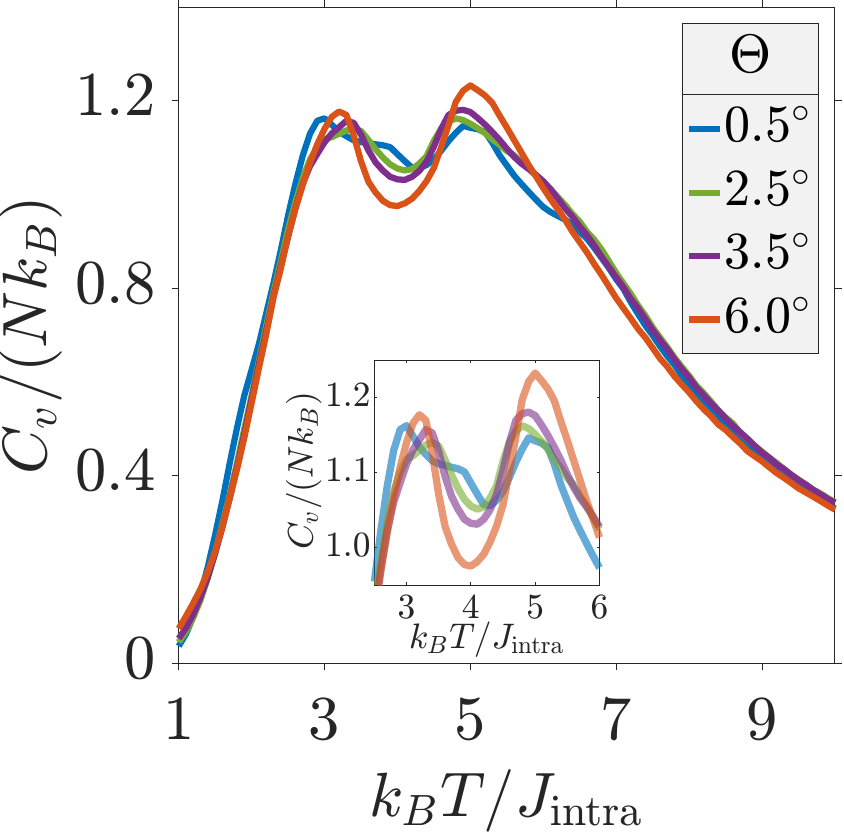}

}\subfloat[$q=6$ (PBC)]{\includegraphics[width=0.263\textwidth]{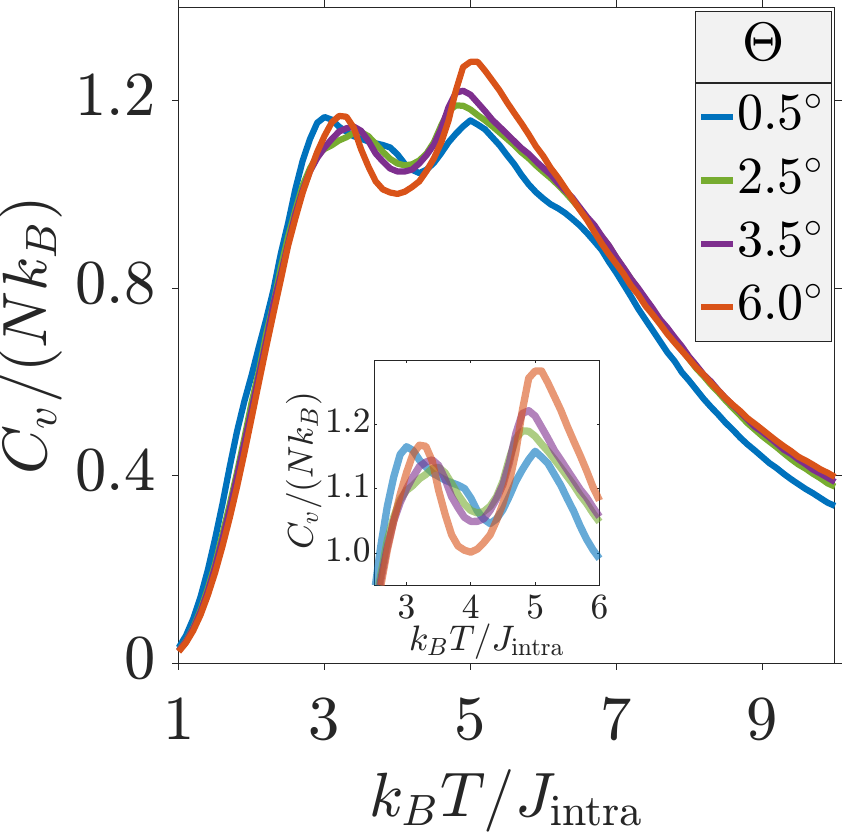}

}\subfloat[$q=6$ (CBC)]{\includegraphics[width=0.263\textwidth]{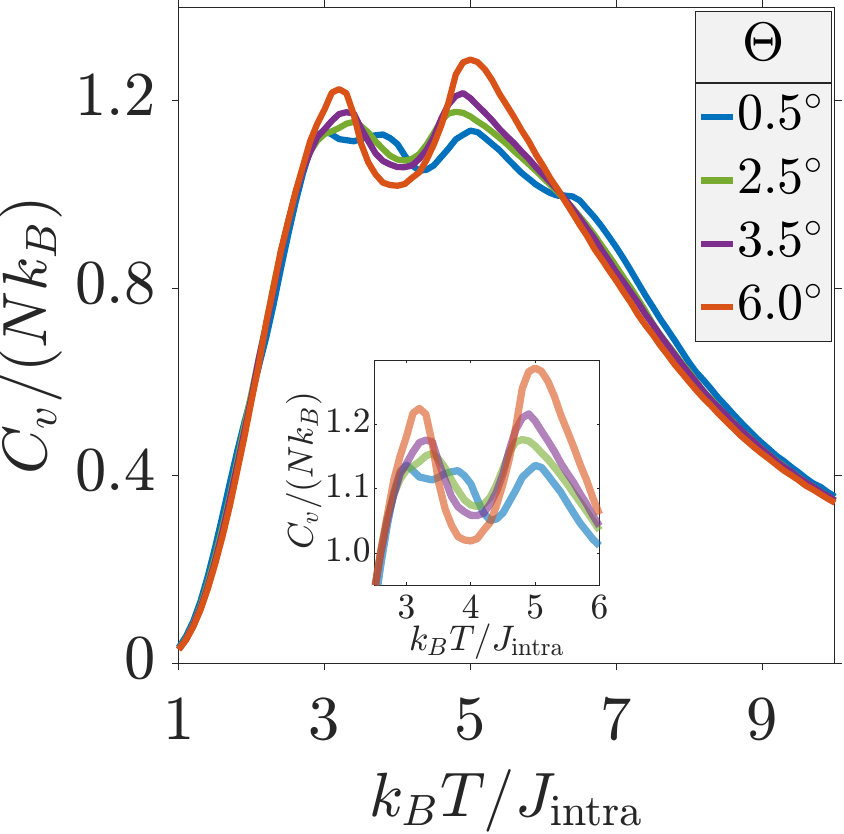}

}

\caption{Specific-heat response for the $q=4$ and $q=6$ cases. Panels (a)--(c)
correspond to $q=4$, and panels (d)--(f) correspond to $q=6$. The
first, second, and third columns correspond to open boundary conditions
(OBC), periodic boundary conditions (PBC), and circular boundary conditions
(CBC), respectively. For all three boundary conditions, the $q=4$
case exhibits the same single-double peak transition as the $q=3$
case, whereas the $q=6$ case, similar to its single-layer counterpart,
does not exhibit such a transition; instead, only the double-peak
profile changes quantitatively with $\Theta$.}\label{fig:ApE}
\end{figure*}

In the main text, we used $q=3$ and $q=5$ as representative cases.
In this section, we show that the twisted bilayer clock model exhibits
thermodynamic anomalies for $q=2$--5, while for $q>5$ the thermodynamic
properties remain qualitatively the same as in their single-layer
counterparts. Here, we provide additional simulation results for $q=4$
to demonstrate the continuity of the thermodynamic anomaly between
$q=3$ and $q=5$, and for $q=6$ as a selected example to confirm
the transition behavior for $q>5$. To further establish the robustness
of the anomaly nature in the bilayer clock model, we also simulate
the system under three different boundary conditions: OBC, PBC, and
CBC.

As shown in Fig.~\ref{fig:ApE}, the first row corresponds to the
$q=4$ twisted bilayer clock model under the three boundary conditions,
while the second row corresponds to the $q=6$ model. For $q=4$,
as $L$ increases, the specific-heat curves exhibit a single-peak
to double-peak to single-peak transition under all boundary conditions,
consistent with the thermodynamic behavior observed for the $q=3$
case and thus confirming the continuity between the $q=3$ and $q=5$
models. In contrast, for $q=6$, no qualitative change occurs in the
specific-heat curves for any of the boundary conditions, and the double-peak
structure is always retained. The variation of $\Theta$ only quantitatively
modifies the profile of this double-peak structure. For $q>6$, the
specific-heat curves exhibit essentially the same thermodynamic behavior
as those for $q=6$ as $\Theta$ varies, with differences only in
the characteristic temperatures and peak magnitudes; these cases are
omitted here for brevity.

In summary, we have demonstrated that the twisted bilayer $q=4$ clock
model exhibits thermodynamic behavior similar to that of the $q=3$
case, confirming the continuity of the anomalous response between
$q=3$ and $q=5$ in the twisted bilayer system. For $q>5$, the system
generally retains its original thermodynamic properties, even when
the twist angle is varied continuously. How to qualitatively modify
the behavior of the $q>5$ bilayer systems, if desired, may require
tuning additional system parameters---a direction that remains open
for future study.

\section{Statistical error analysis}

\begin{figure*}
\centering
\subfloat[$q=3$]{\includegraphics[width=0.245\textwidth]{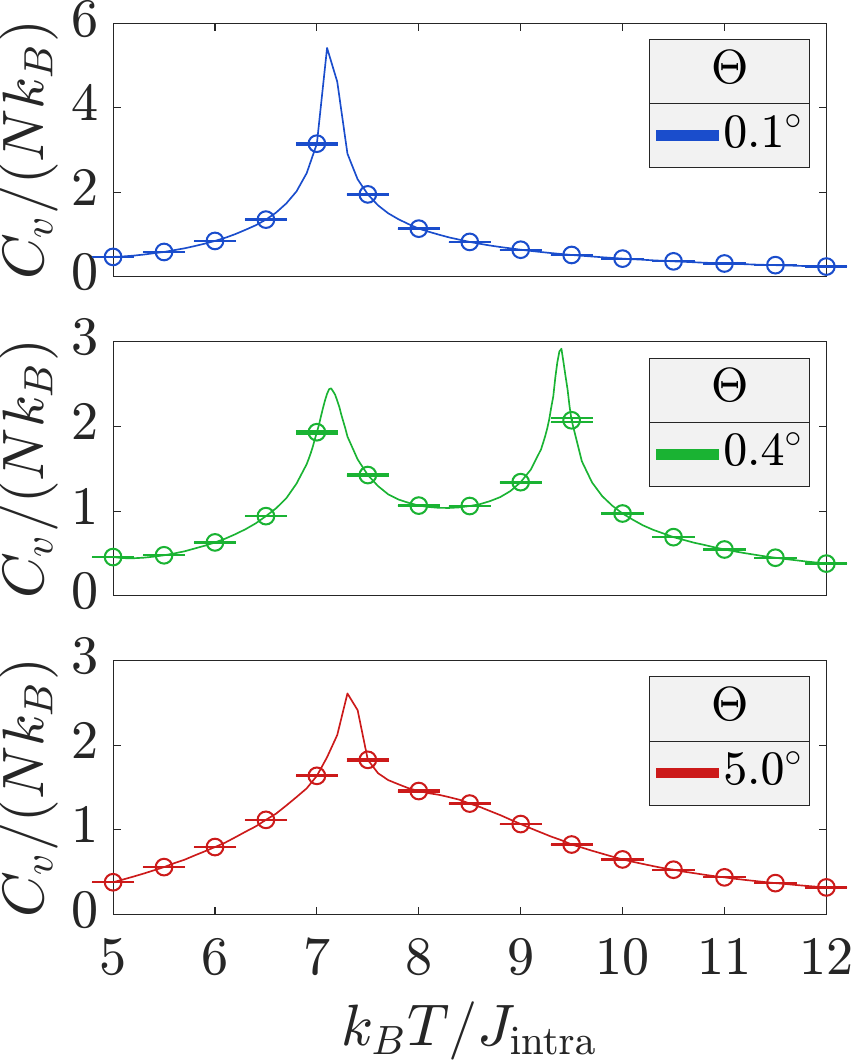}

}\subfloat[$q=4$]{\includegraphics[width=0.245\textwidth]{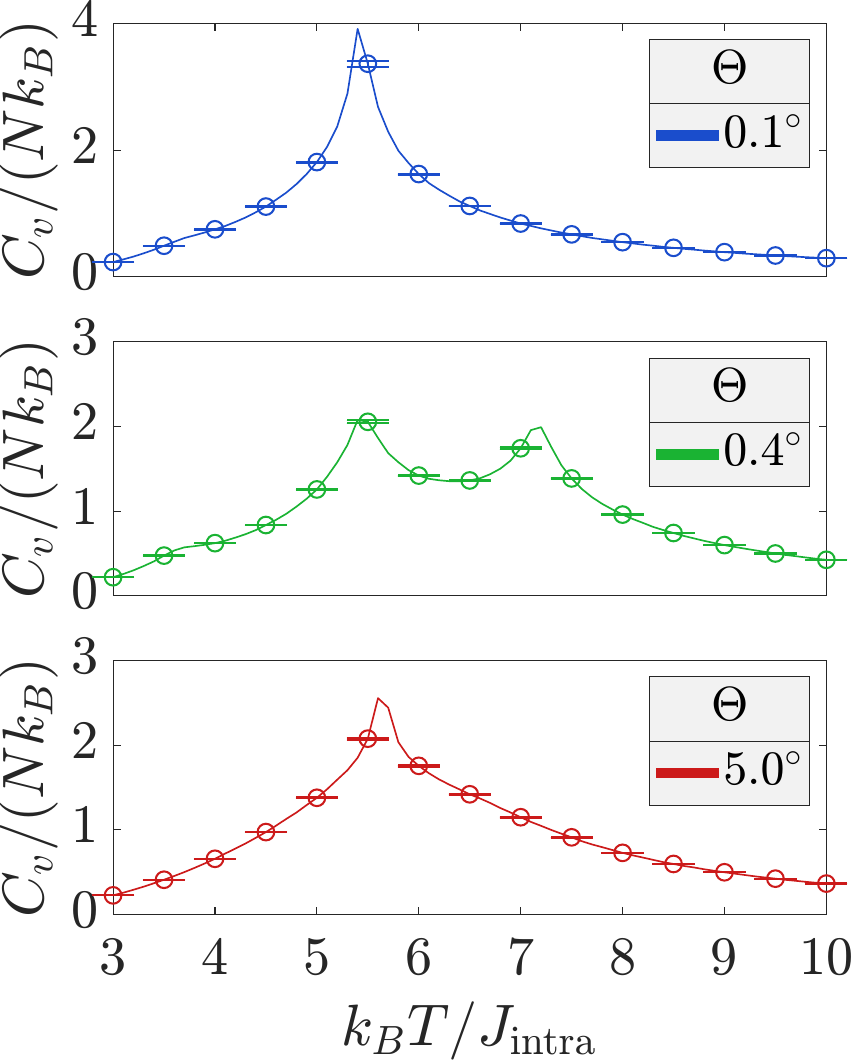}

}\subfloat[$q=5$]{\includegraphics[width=0.245\textwidth]{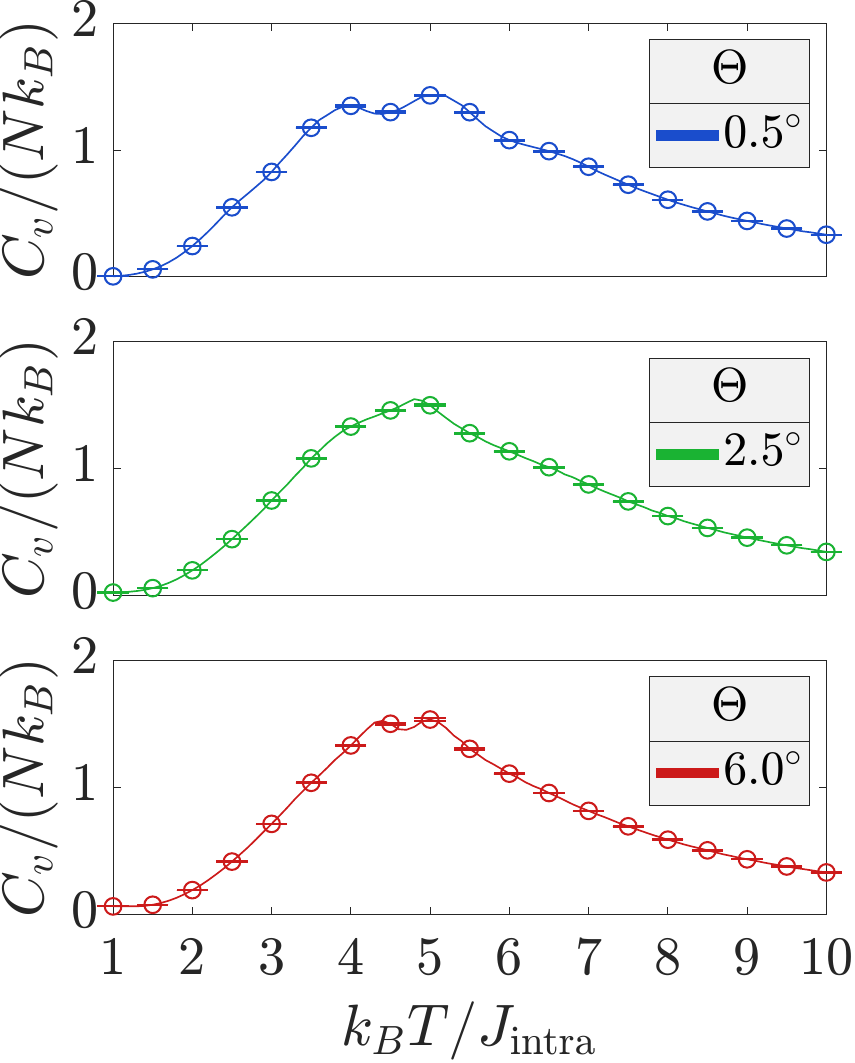}

}\subfloat[$q=6$]{\includegraphics[width=0.245\textwidth]{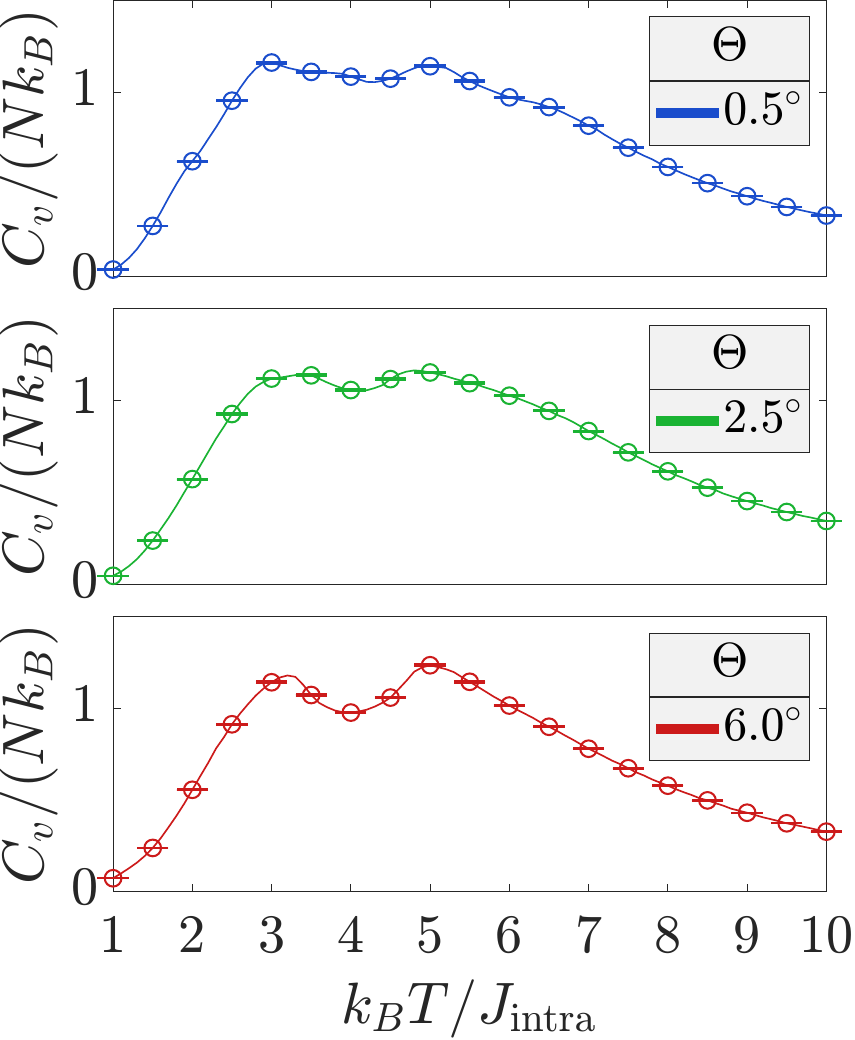}

}

\caption{Error bars in the Monte Carlo simulations. Panels (a)--(d) correspond
to $q=3,4,5$, and 6, respectively. Each panel shows three specific-heat
curves for different twist angles $\Theta$. The simulations are performed
with open boundary conditions (OBC). The error bars are generally
smaller than the symbol size. They are nearly zero in the low- and
high-temperature regimes (approximately $10^{-5}$--$10^{-3}$) and
only slightly larger near the phase-transition region (approximately
0.01--0.03), indicating well-converged data.}\label{fig:ApF}
\end{figure*}

To maintain visual clarity and to highlight the evolution of the specific-heat
curves, we omitted error bars from the main text. Here, we present
a subset of curves for $q=3$--6 with error bars included.

As shown in Fig.~\ref{fig:ApF}, we select three representative specific-heat
curves for each $q$-state clock model under OBC, and the corresponding
error bars are displayed at evenly spaced temperature points using
the same colors as the curves. It is evident from the figure that
the error bars in every panel are very small---smaller than the symbols
marking the data points---regardless of $q$, $\Theta$, or temperature.
The upper and lower caps of each error bar nearly overlap, and the
stem length is negligible compared with the specific-heat magnitude.
For a given specific-heat curve, the error bars in the low- and high-temperature
regimes are nearly zero (approximately $10^{-5}$--$10^{-3}$), whereas
those in the transition-temperature region are slightly larger (approximately
0.01--0.03). This is expected because the system undergoes critical
slowing down near the phase transition, requiring more Monte Carlo
steps to reach equilibrium. Overall, the small error bars confirm
that our Monte Carlo results are statistically reliable, and that
the specific-heat curves faithfully represent the thermodynamic behavior
of the twisted bilayer clock system and can serve as a reliable guide
for future experimental studies.

\bibliographystyle{apsrev4-2}
\bibliography{refs}

\end{document}